\documentclass[usenatbib]{mn2e}

\input{psfig.sty}
\usepackage{graphicx}

\def\beq{\begin{equation}}
\def\enq{\end{equation}}
\def\ms{${\rm ~M}_{\odot}$}

\title{Evidence for a magnetic neutron star in high mass X-ray binary 4U 2206+54 with INTEGRAL/IBIS observations}

\author[Wei Wang]{Wei Wang\thanks{E-mail: wangwei@bao.ac.cn} \\
National Astronomical Observatories, Chinese Academy of Sciences,
Beijing 100012, China}

\begin{document}
\maketitle

\begin{abstract}

The hard X-ray source 4U 2206+54 is a peculiar high mass X-ray
binary with a main-sequence donor star. Recent X-ray observations
suggested that the compact object in 4U 2206+54 may be a neutron
star. The X-ray emission comes from the accretion of stellar winds
from the massive donor stars, and variability of luminosity may be
due to the changes of its orbit phase. To further reveal the
nature of compact object, we studied 4U 2206+54 with INTEGRAL/IBIS
observations in two years, and found that in most time, 4U 2206+54
undergone a quiescent state and sometime an active state. In the
quiescent state the spectrum can be fitted by a power-law model of
$\Gamma\sim 2.1$ with a hard X-ray luminosity of $\sim 5\times
10^{34}$ erg s$^{-1}$ (20-- 100 keV). While in the active state,
the 20-- 100 keV hard X-ray luminosity reaches $\sim 2\times
10^{35}$ erg s$^{-1}$ and the spectrum is fitted by a thermal
bremmstrahlung model of $kT\sim 43$ keV plus two cyclotron
absorption lines at $\sim$ 30 and 60 keV. Then we derived a
magnetic field of 3.3$\times 10^{12}$ G for the compact object in
4U 2206+54. During the active state, we found a pulsation period
of $\sim$ 5400 s in the light curve of 4U 2206+54. So the compact
object in 4U 2206+54 should be a magnetic neutron star with a slow
pulsation. Cyclotron absorption lines detected in the active state
and non-detection in the quiescent state suggested that two
different accretion states have possible different hard X-ray
emission regions: surface of neutron star in the active state; the
magnetic-accretion pressure equivalent point in the quiescent
state. The re-analysis of the RXTE/ASM light curve found the
modulation periods at $\sim 9.56$ days and 19.11 days, and the
orbit period of 4U 2206+54 should be 19.11 days.

\end{abstract}


\section{INTRODUCTION}

The X-ray source 4U 2206+54 has been studied with numerous ground
and space-based observations, but the nature of this source and
origins of its variability patten are still unclear. 4U 2206+54
was identified with an optical counterpart BD +53 2790 by Steiner
et al. (1984) which was initially classified as a Be star.
However, in optical and UV bands the emission spectrum is complex,
in particular the behavior of the H$\alpha$ emission line,
suggesting that this star is an O9.5V star (Negueruela \& Reig
2001; Rib\'o et al. 2006;  Blay et al. 2006) with a high He
abundance (Blay et al. 2006).

X-ray monitoring of 4U 2206+54 by RXTE suggested a modulation
period of 9.6 days (Corbet \& Peele 2001) which may be an orbit
period, but this period disappeared with recent SWIFT/BAT
observations (Corbet et al. 2007) which instead found a modulation
of 19.25 days consistent with twice the 9.6-day period. Since
there is no circumstellar disc around the donor of 4U 2206+54, the
material needed for accretion and production of X-rays should come
from the stellar wind. The observed X-ray luminosity of 4U 2206+54
varies from 10$^{33}-10^{35}$ erg s$^{-1}$ from the RXTE,
BeppoSAX, SWIFT and INTEGRAL light curves between 1996 and 2005
(Torrejon et al. 2004; Masetti et al. 2004; Blay et al. 2005;
Corbet et al. 2007). Rib\'o et al. (2006) found a low wind
terminal velocity of $\sim 350$ km s$^{-1}$ in 4U 2206+54. With
such a low wind terminal velocity, and assuming an eccentric
orbit, one could reproduce the X-ray luminosity and orbit
variability of the system (Rib\'o et al. 2006).

The nature of the compact object in 4U 2206+54 has been in dispute
for a long time (Negueruela \& Reig 2001; Corbet \& Peele 2001).
Broad band X-ray observations and radio studies on 4U 2206+54
favored the presence of a neutron star (Torrejon et al. 2004; Blay
et al. 2005). Recent reports on the possible detection of electron
cyclotron resonant absorption line at $\sim 30$ keV suggested a
magnetic field of $\sim 3\times 10^{12}$ G by different
observations of RXTE, BeppoSAX and INTEGRAL (Torrejon et al. 2004;
Masetti et al. 2004; Blay et al. 2005). Non-detection of pulsation
in 4U 2206+54 was always used as a doubt on the neutron star
scenario. Recently, a possible 5500-s pulsation period in the
light curve of 4U 2206+54 was discovered with the RXTE
observations (Reig et al. 2009). If it is true, the compact object
in 4U 2206+54 will be a neutron star with a spin-period of 5500
seconds.

In this work, we will study hard X-ray characteristics of 4U
2206+54 with the INTEGRAL/IBIS observations from 2003 May to 2005
December. Then we can show the hard X-ray spectral properties both
in quiescent states (average X-ray luminosity $\sim 10^{34}$ erg
s$^{-1}$) and active states ($>10^{35}$ erg s$^{-1}$). Two
cyclotron resonant absorption lines (fundamental and first
harmonic) were found in the spectrum of the active states during
2005 Dec, while in the quiescent states, no cyclotron line
features are detected. The existence of cyclotron absorption lines
suggested a strong magnetic field in 4U 2206+54, and the high
energy emission in active states would mainly come from
magnetosphere near the neutron star. Therefore, it is possible
that we could search for a pulsation period from the light curve
of 4U 2206+54 in active states during 2005 Dec with INTEGRAL/IBIS
observations. Finally, in order to confirm the existence of the
orbit period in X-ray binary 4U 2206+54, we also re-analyzed the
archival data of the All Sky Monitor (ASM) aboard RXTE from 1997
to 2008 to search for the orbit period.

\begin{table*}

\caption{INTEGRAL/IBIS observations of the field around 4U
2206+54. The time intervals of observations in the revolution
number and the corresponding dates, the corrected on-source
exposure time are listed. And mean count rate and the detection
significance level value in the energy range of 20 -- 60 keV were
also shown. }

\begin{center}
\scriptsize
\begin{tabular}{l c c c l}

\hline \hline Rev. Num. & Date  & On-source time (ks) & Mean count rate s $^{-1}$ & Detection level \\
\hline 67 & 2003 May 02--04 & 11 & 5.6$\pm 0.3$& 17$\sigma$ \\
87 & 2003 Jul 01 -- 03 & 9 & $6.6\pm 0.4$ & 16$\sigma$ \\
142 -- 148 & 2003 Dec 13 -- 31 & 92 & 0.6$\pm 0.1$ & 5.6$\sigma$ \\
161 --162 & 2004 Feb 07 -- 12 & 33 & $0.7\pm 0.2$ & 5.1$\sigma$ \\
262 -- 264 & 2004 Dec 05 -- 13 & 53 & 0.6$\pm 0.1$ & 5.5$\sigma$ \\
384 --386 & 2005 Dec 05 -- 13 & 68 & 3.3$\pm 0.2$ & 20$\sigma$ \\
\hline
\end{tabular}
\end{center}

\end{table*}

\begin{figure}
\centering
\includegraphics[angle=0,width=8.5cm]{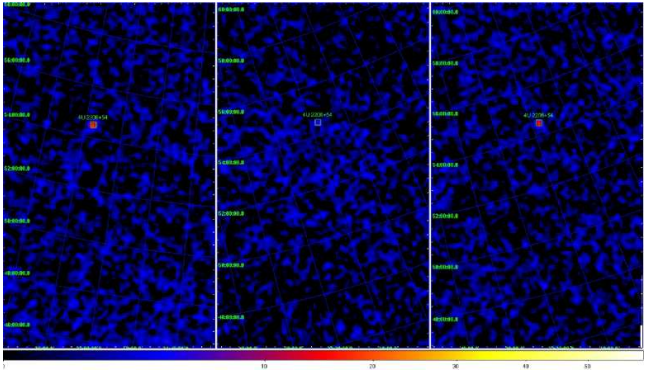}
\caption{Significance mosaic maps around 4U 2206+54 in Equatorial
J2000 coordinates as seen with INTEGRAL/IBIS in the energy range
of 20 - 60 keV during three observational time intervals (from
left to right, detection significance level also noted): 2003 May
2 ($\sim 17\sigma$); 2003 Dec 13 to 31 ($\sim 5\sigma$); 2005 Dec
5 to 12 ($\sim 20\sigma$). False color representation of
significance is displayed on a logarithmic scale.  }
\end{figure}

\begin{figure}
\centering
\includegraphics[angle=0,width=7cm]{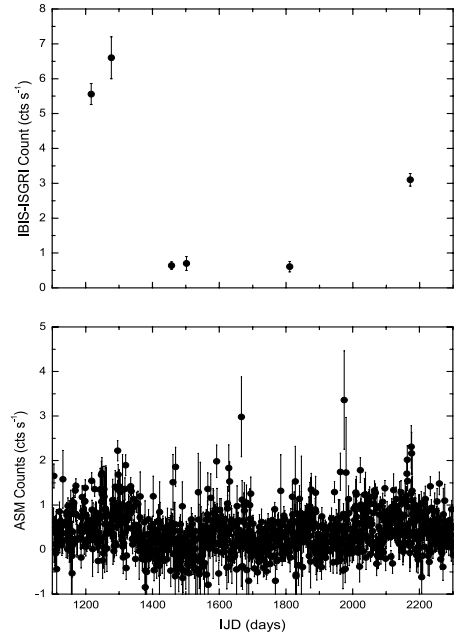}
\caption{{\bf Top:} The light curve of X-ray binary 4U 2206+54 in
the energy range of 20-- 60 keV from IBIS observations from 2003
May to 2005 Dec. The time unit is the INTEGRAL Julian date (IJD)
starting at 1 Jan 2000.  {\bf Bottom:} The light curve of X-ray
binary 4U 2206+54 in the energy range of 1.5-- 12 keV from 2003
May to 2005 Dec from the RXTE/ASM data.  }
\end{figure}

\section{Observations}

The INTErnational Gamma-Ray Astrophysics Laboratory (INTEGRAL,
Winkler et al. 2003) is ESA's currently operational space-based
hard X-ray/soft gamma-ray telescope. There are two main
instruments aboard INTEGRAL, the imager IBIS (Ubertini et al.
2003) and the spectrometer SPI (Vedrenne et al. 2003),
supplemented by two X-ray monitors JEM-X (Lund et al. 2003) and an
optical monitor OMC (Mas-Hesse et al. 2003). All four instruments
are co-aligned, allowing simultaneous observations in a wide
energy range.

4U 2206+54 was observed during the INTEGRAL surveys of the Cygnus
and Cassiopeia regions. 4U 2206+54 cannot be detected by JEM-X
which has a smaller field of view (FOV), and SPI has the largest
FOV but larger uncertainties compared with IBIS, so we only use
IBIS data for the analysis in this work. The data were collected
with the low-energy array called ISGRI (INTEGRAL Soft Gamma-Ray
Imager) which consists of a pixellated 128$\times 128$ CdTe
solid-state detector that views the sky through a coded aperture
mask (Lebrun et al. 2003). IBIS/ISGRI has a 12' (FWHM) angular
resolution and arcmin source location accuracy in the energy band
of 15 -- 200 keV. In Table 1, we summarize the INTEGRAL
revolutions in our analysis. We used the archival data which are
available from the INTEGRAL Science Data Center (ISDC).

The analysis was done with the standard INTEGRAL off-line
scientific analysis (OSA, Goldwurn et al. 2003) software, ver.
7.0. In this latest version, the energy correction has changed and
new calibration laws have been used for the IBIS/ISGRI analysis.
The drift of the energy calibration gain and offset with
activation and time are much better calibrated at intermediate
energies (around 50 keV). This version results in more constant
Crab light curve. Individual pointings processed with OSA 7.0 were
mosaicked to create sky images according to the methods and
processes described in Bird et al (2007). And we have used the 20
-- 60 keV band for source detection and to quote fluxes.

We displayed the significance mosaic maps around 4U 2206+64 as
seen with INTEGRAL/IBIS in the energy range of 20 -- 60 keV during
three observational time intervals in Figure 1 and the light curve
of 4U 2206+54 from 2003 May to 2005 December obtained by both IBIS
observations in Figure 2 (top). We also presented the light curve
of 4U 2206+54 in the energy band of 1.5 --12 keV from the RXTE/ASM
long-term monitoring for a comparison (Figure 2 bottom). The count
rate from 4U 2206+54 in the energy range of 20 --60 keV varied in
more than two years: $\sim 6$ cts s$^{-1}$ on 2003 May 2 and 2003
July 2; and undergoing a quiescent state, $\sim 0.7$ cts s$^{-1}$
around 2003 Dec, 2004 Feb and 2004 Dec; again an active state,
$\sim 3.3$ cts s$^{-1}$ around 2005 Dec 5 --13 (also see Table 1).

\section{Spectral analysis}

The spectral results on the source 4U 2206+54 during the
revolutions 67 and 87 with INTEGRAL/IBIS data have been presented
in Blay et al. (2005). They reported a possible cyclotron
absorption line around 32 keV. Here we will not show the spectral
results during these time intervals again, but carry out spectral
analyses on the data during the quiescent state and the active
state around Dec 2005 respectively. The spectral analysis software
package used is XSPEC 12.4.0x (Arnaud 1996).

For the quiescent state, we summed up images from the INTEGRAL
resolutions 142 -- 148, 161 --162 and 262 -- 264 (see Table 1) to
improve the detection significance level, and then derived the
spectrum which was shown in Figure 3. The 20 -- 150 keV spectrum
can be fitted with a power-law model of a photon index $\Gamma\sim
2.1 \pm 0.3$. The obtained flux in the quiescent state from 20
--100 keV is $(4.9\pm 0.9)\times 10^{-11}$ erg cm$^{-2}$ s$^{-1}$,
corresponding to a hard X-ray luminosity of $\sim 5.3\times
10^{34}$ erg s$^{-1}$ assuming a distance of $\sim 3$ kpc (Blay et
al. 2006). No significant cyclotron absorption features were found
in quiescence. If assuming the presence of the cyclotron
absorption lines at $\sim 30$ keV (Blay et al. 2005) and $\sim 60$
keV (see below), we derived the upper limits on the equivalent
width (EW) of the absorption lines: $<1.6$ keV at 30 keV
(2$\sigma$) and $<2.4$ keV at 60 keV (2$\sigma$).

The same spectral analysis was carried out during the active state
of 4U 2206+54 from Dec 5 to 12, 2005. The derived spectrum was
displayed in Figure 4. In the active state, a simple power-law
model cannot fit the spectrum, we used a thermal bremsstrahlung
model to fit it. However, possible absorption features around 30
and 60 keV cannot be fitted (also seen in the residuals in Figure
4 top panel). Therefore, we used the thermal bremsstrahlung model
to fit the continuum, and added the cyclotron resonant absorption
model by using the XSPEC cyclabs to the continuum fit (see Figure
3). In this case, we found a thermal bremsstrahlung model of $kT
\sim 43.1\pm 2.0$ keV plus two electron cyclotron resonant
absorption lines at $\sim 29.6\pm 2.8$ keV (F-test probability:
2.9$\times 10^{-6}$) with a FWHM of $\sim 1.8\pm 0.5$ keV and
$\sim 59.5\pm 2.1$ keV (F-test probability: 4.2$\times 10^{-10}$)
with a FWHM of $\sim 3.9\pm 0.9$ keV (reduced $\chi^2 \sim 0.95$,
6 {\em d.o.f}). The derived EW is $\sim 2.2\pm 0.7$ keV at $\sim
30$ keV and $\sim 9.1\pm 1.7$ keV at $\sim 60$ keV separately.

The obtained flux from the 20 -- 100 keV continuum during the
active state in 4U 2206+54 is $\sim (2.1\pm 0.1)\times 10^{-10}$
erg cm$^{-2}$ s$^{-1}$, corresponding to a hard X-ray luminosity
of $\sim 2.3\times 10^{35}$ erg s$^{-1}$.

\begin{figure}
\centering
\includegraphics[angle=-90,width=9cm]{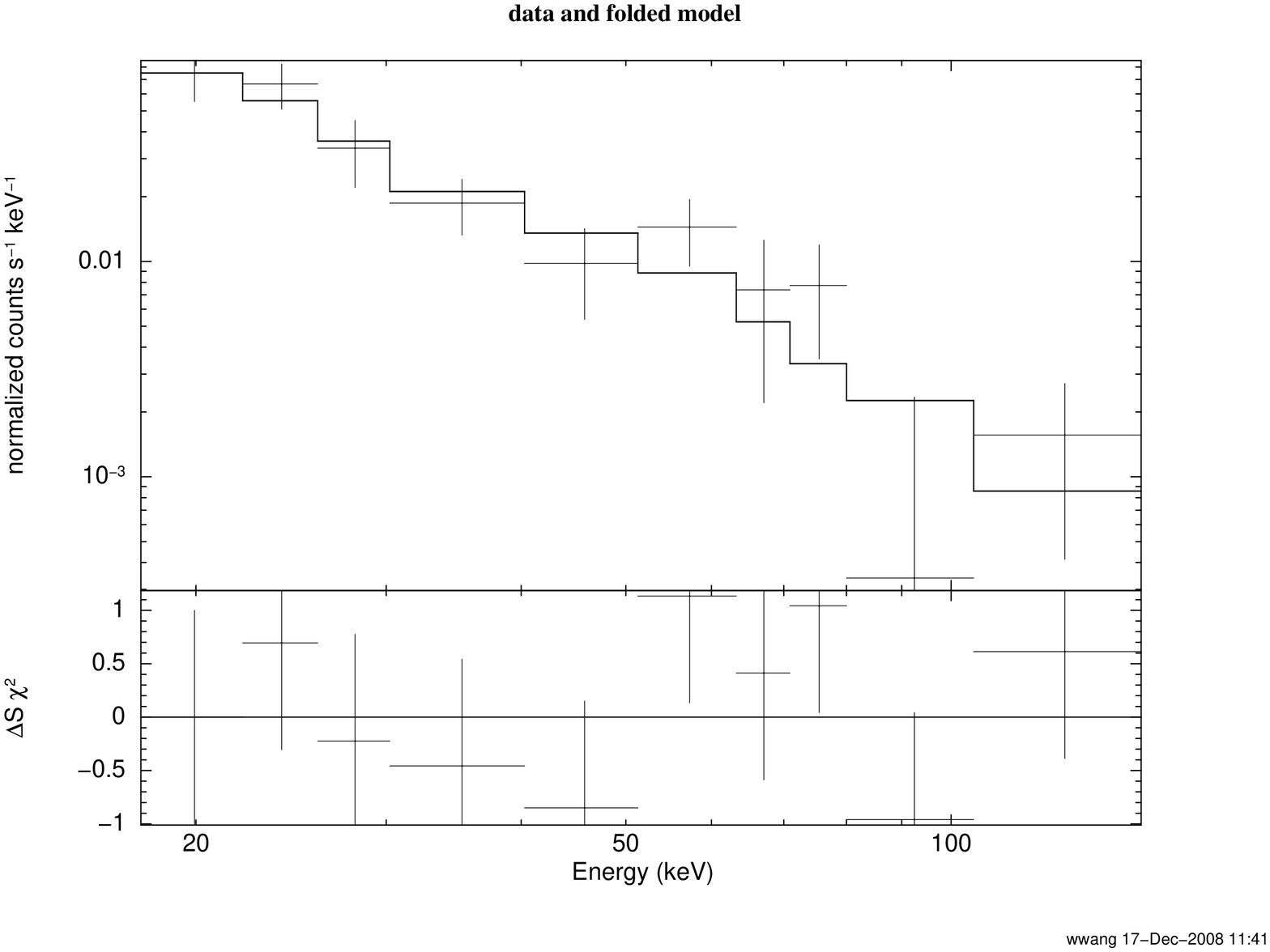}
\caption{The spectrum of 4U 2206+54 in the quiescent state which
was derived from the summed mosaic image from the INTEGRAL
resolutions 142 -- 148, 161 --162 and 262 -- 264. The spectrum can
be described by a power-law model with a photon index of
$\Gamma\sim 2.1\pm 0.3$ (reduced $\chi^2 \sim 0.69$, 8 {\em
d.o.f}). }

\end{figure}

\begin{figure}
\centering
\includegraphics[angle=-90,width=9cm]{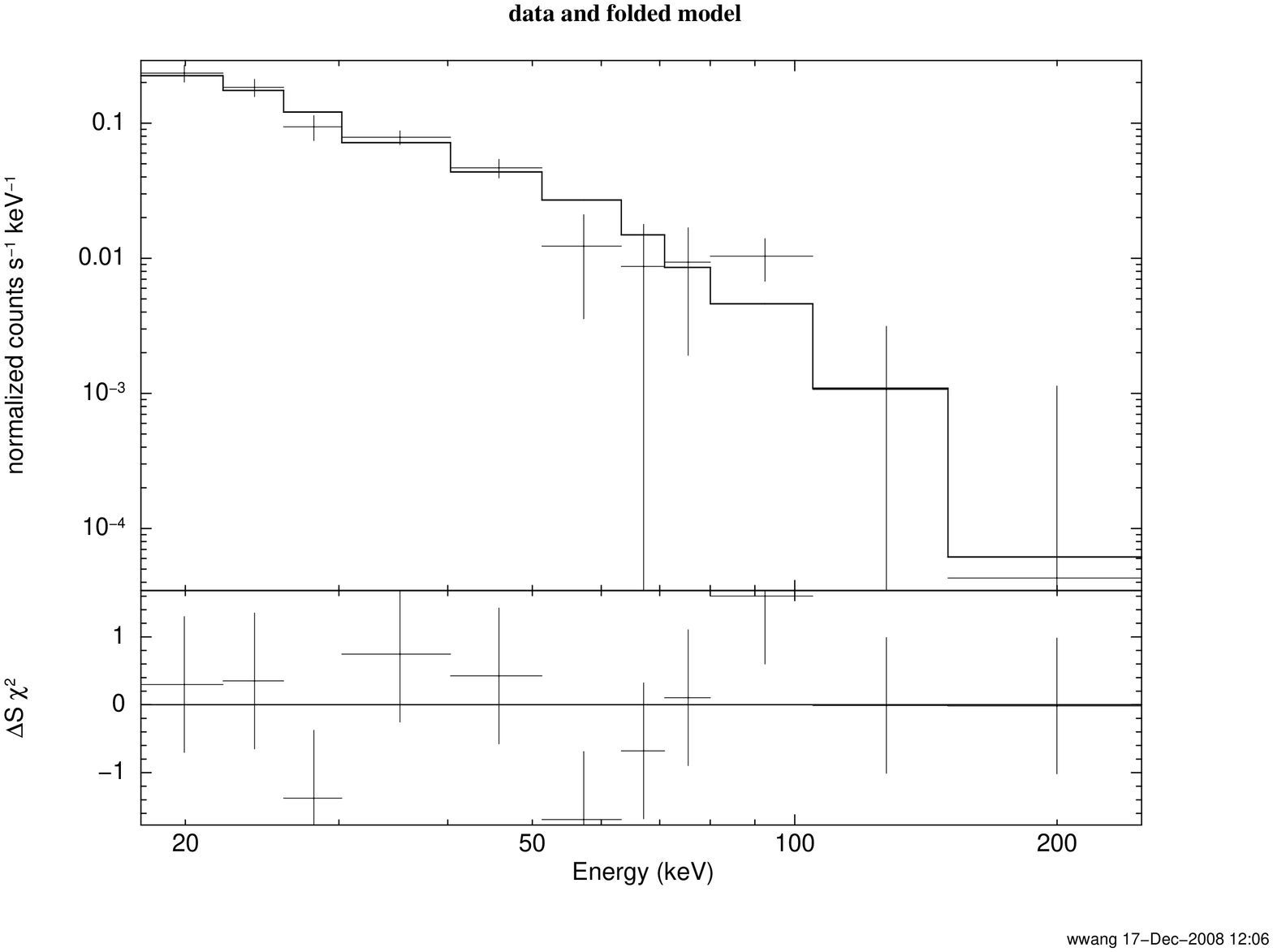}
\includegraphics[angle=-90,width=9cm]{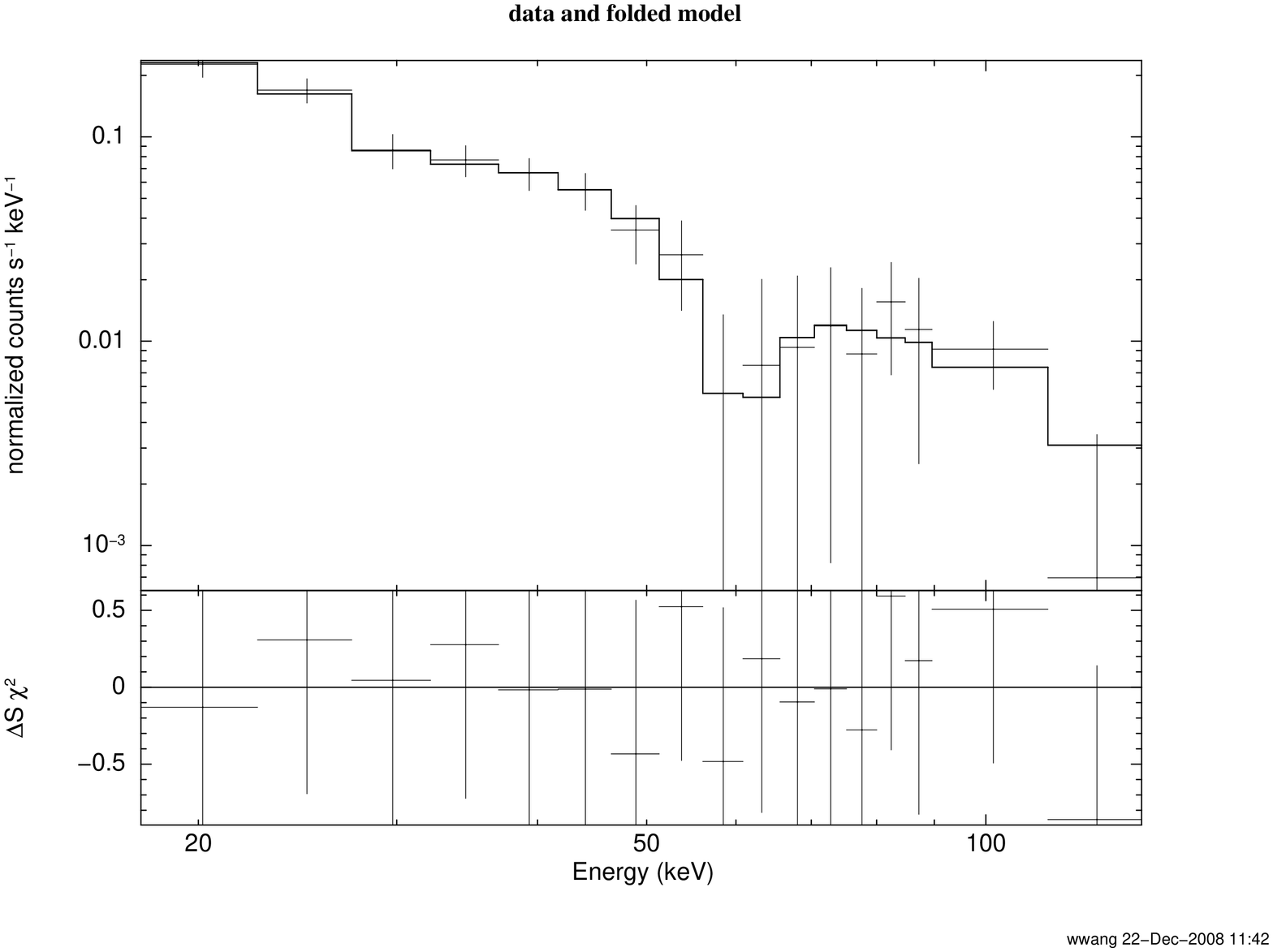}
\caption{The spectrum of 4U 2206+54 in the active state from Dec 5
to 12, 2005. {\bf Top} the spectrum only fitted with a thermal
bremsstrahlung model of $kT \sim 37.9\pm 1.2$ keV (reduced $\chi^2
\sim 1.68$, 9 d.o.f). The possible absorption features around 30
keV and 60 keV cannot be fitted well. {\bf Bottom} the spectrum
(re-binned for better line fits) can be fitted by a thermal
bremsstrahlung model of $kT \sim 43.1\pm 2.0$ keV plus two
cyclotron resonant absorption lines at $\sim 29.6\pm 2.8$ keV and
$\sim 59.5\pm 2.1$ keV (reduced $\chi^2 \sim 0.95$, 6 {\em
d.o.f}). }

\end{figure}

It is the first time that we detected both the fundamental and
second harmonic of the cyclotron resonant absorption lines in the
X-ray binary 4U 2206+54. The detection of the line feature at
$\sim 30$ keV also confirmed the previous reports from different
measurements (Torrejon et al. 2004; Massetti et al. 2004; Blay et
al. 2005). Discovery of the cyclotron resonant absorption lines at
$\sim 30$ and 60 keV strongly suggested a magnetic neutron star
located in the binary 4U 2206+54.

We can calculated the value of the magnetic field of the neutron
star in 4U 2206+54 by using the formula \beq [B/10^{12}{\rm
G}]=[E_{\rm cycl}/11.6{\rm keV}](1+z), \enq where $E_{\rm cycl}$
is the energy of the fundamental line, here $E_{\rm cycl}=29.6$
keV, and $z$ is the gravitational redshift near the surface of the
neutron star. For a canonical neutron star of 1.4 \ms with a
radius of 10 km, we can take $z\sim 0.3$ (Kreykenbohm et al.
2004). So we obtain a magnetic field of $3.3\times 10^{12}$ G for
the neutron star in 4U 2206+54, and this value is still in
agreement with those in the previous studies (Torrejon et al.
2004; Massetti et al. 2004; Blay et al. 2005).

\section{Searching for the pulsation and orbit periods}

The high mass X-ray binary 4U 2206+54 shows a variable X-ray light
curve. Intensive searches for the pulsation period of 4U 2206+54
have been performed with EXOSAT (Corbet \& Peele 2001), RXTE
(Negueruela \& Reig 2001; Torrejon et al. 2004; Corbet et al.
2007), BeppoSAX (Torrejon et al. 2004; Massetti et al. 2004) and
INTEGRAL (Blay et al. 2005). However, these studies suggested the
lack of the X-ray pulsation on timescales from $\sim 1$ ms to $1$
hr. Recently, Reig et al. (2009) use the new RXTE observational
data to search for the pulsation period longer than 1 hr, and
discovered a possible 5500-s pulsation period in the light curve
of 4U 2206+54. In addition, the early RXTE/ASM data suggested a
modulation period of $\sim 9.6$ days (Corbet \& Peele 2001), but
the SWIFT/BAT data reported a period of $\sim 19.25$ days (Corbet
et al. 2007), so that Corbet et al. (2007) suggested the orbit
period should be 19.25 days instead of 9.6 days. Here we will
check these period reports with recent data of INTEGRAL/IBIS and
RXTE/ASM.

We have reported the cyclotron resonant absorption line features
during the active state in 4U 2206+54, suggesting that during the
active state, the X-ray emission mainly coming from the surface of
the neutron star. So we try to search for a pulsation period using
the light curve data during the active state. We applied the FFT
to the observational intervals from Dec 5 to 12, 2005. The power
spectrum was shown in Figure 5, binned at 100 s intervals. A
significant period signal was found at $\sim 5400^{+300}_{-200}$ s
(Figure 5 right). This period is a little lower than that reported
by Reig et al. (2009), but still consistent considering the error
bar range. Thus, 4U 2206+54 should be a X-ray pulsar with a very
slow pulsation. The folded light curve of 4U 2206+54 at a
pulsation period (5400 s) is also shown in Figure 6. A possible
double main peak feature appears in the pulse profile, one at
$\sim 0.2$ and the other around 0.6 -- 0.9.

Since the available database during the active state is limited to
less than 7 days, we cannot search for the longer modulation
period (like orbit period, possible $\sim$ 9.6 d or 19.2 d) with
the INTEGRAL database during the active state of 4U 2206+54. We
used archival RXTE/ASM data with observations from 1997 -- 2008 to
search for the orbit period in 4U 2206+54. Firstly, we averaged
the ASM 1.5 -- 12 keV light curves into 1 hour bins. After
subtracting the mean value of all inhabited bins from each
inhabited bin, we take an FFT to the light curve. The power
spectrum is shown in Figure 7. In the top panel of Fig. 7, we have
shown a power spectrum in a wider frequency range, and no
significant signal was found in the high frequency range ($<100$
hours), while two significant peaks were detected in the lower
frequency range. Then in the bottom panel, from the zoom-in of the
low frequency band, we determined two modulation periods at
9.56$\pm 0.03$ days and $19.11\pm 0.07$ days (twice the former
one). So we confirm the existence of the possible orbit periods
reported by both the early RXTE/ASM data (Corbet \& Peele 2001)
and the recent SWIFT/BAT observations (Corbet et al. 2007). In
Figure 8, the folded light curves of 4U 2206+54 using the RXTE/ASM
data at two modulation periods at 9.56 day and 19.11 day were also
displayed separately. A possible peak at $\sim 33.2$ days also
appeared in the power spectrum though not significantly yet, which
may be just a noise signal.

\section{Conclusion and discussion}

\begin{figure*}
\centering
\includegraphics[angle=0,width=8cm]{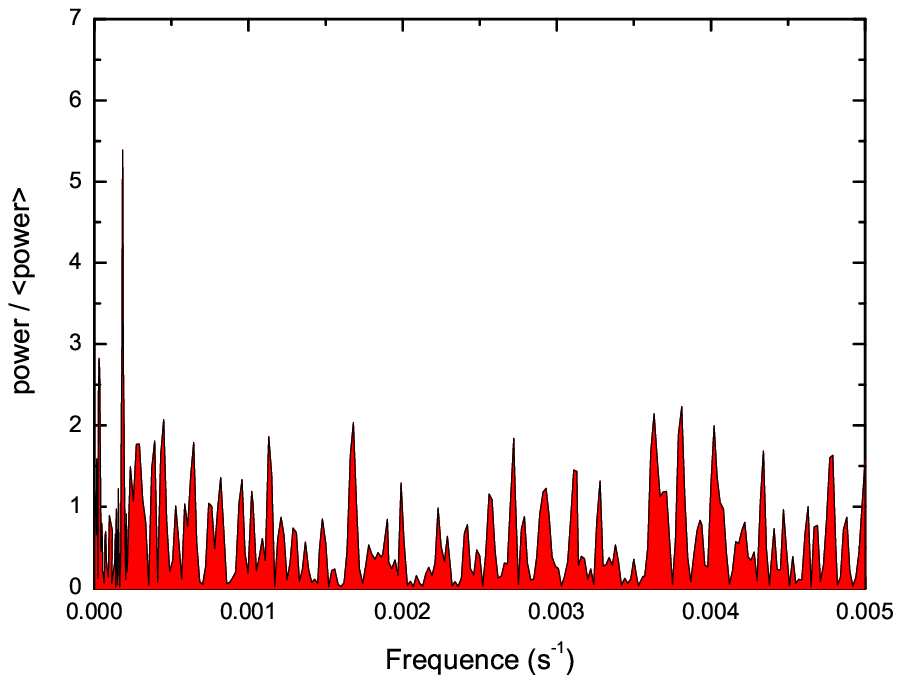}
\includegraphics[angle=0,width=8cm]{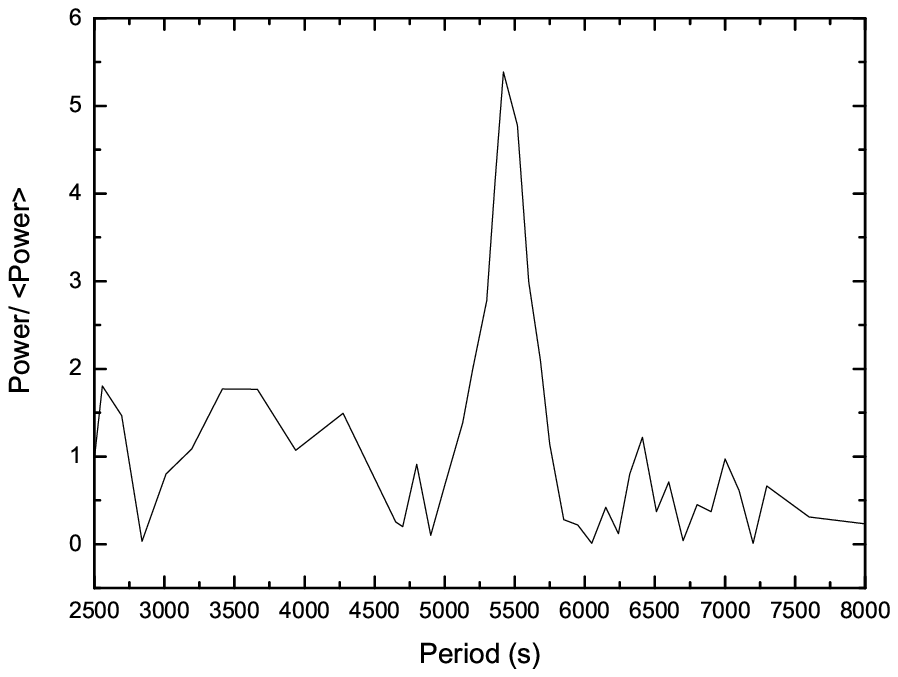}
\caption{{\bf Left} Power spectrum of the INTEGRAL/IBIS light
curve of 4U 2206+54 during the active state. The significant peak
appears around 5000 s. The unit is the resulting power in each
frequency bin divided by the average power over the whole
frequency range (same in Fig. 6). {\bf Right} The zoom-in of the
power spectrum around 5000 s. We determined a period of $\sim
5400^{+300}_{-200}$ s. }
\end{figure*}

\begin{figure}
\centering
\includegraphics[angle=0,width=7cm]{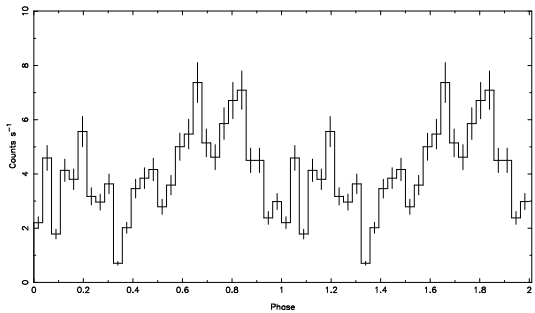}
\caption{The IBIS-ISGRI background subtracted light curve (20 --
60 keV) of 4U 2206+54 folded at a pulsation period (5400 s). The
pulse profile is repeated once for clarity. }
\end{figure}

In this paper, we studied the hard X-ray emission of the high mass
X-ray binary 4U 2206+54 with INTEGRAL/IBIS observations to reveal
the nature of the compact object. In hard X-ray bands, 4U 2206+54
undergone both the quiescent and active states. In the quiescent
state, the 20 -- 150 keV spectrum of 4U 2206+54 can be fitted by a
power-law model with $\Gamma \sim 2.1$, and no cyclotron
absorption line features are found. While, in the active state,
two cyclotron absorption lines at $\sim 30$ keV and 60 keV are
discovered in the hard X-ray spectrum. With known the cyclotron
resonant absorption energy, we determined the magnetic field of
$\sim 3.3\times 10^{12}$ G for the compact object in 4U 2206+54.
Analysis of the X-ray light curve during the active state, we also
found a modulation period of $\sim 5400$ s which would be a
pulsation period for a neutron star. Thus we identified the
compact object in 4U 2206+54 as a magnetic neutron star with a
very slow pulsation.

\begin{figure*}
\centering
\includegraphics[angle=0,width=7cm]{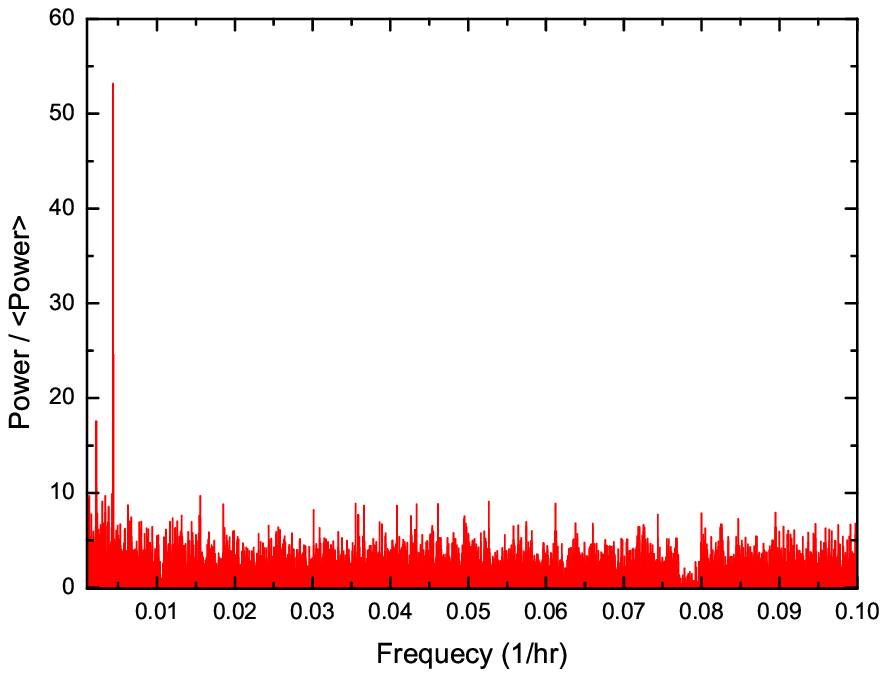}
\includegraphics[angle=0,width=7cm]{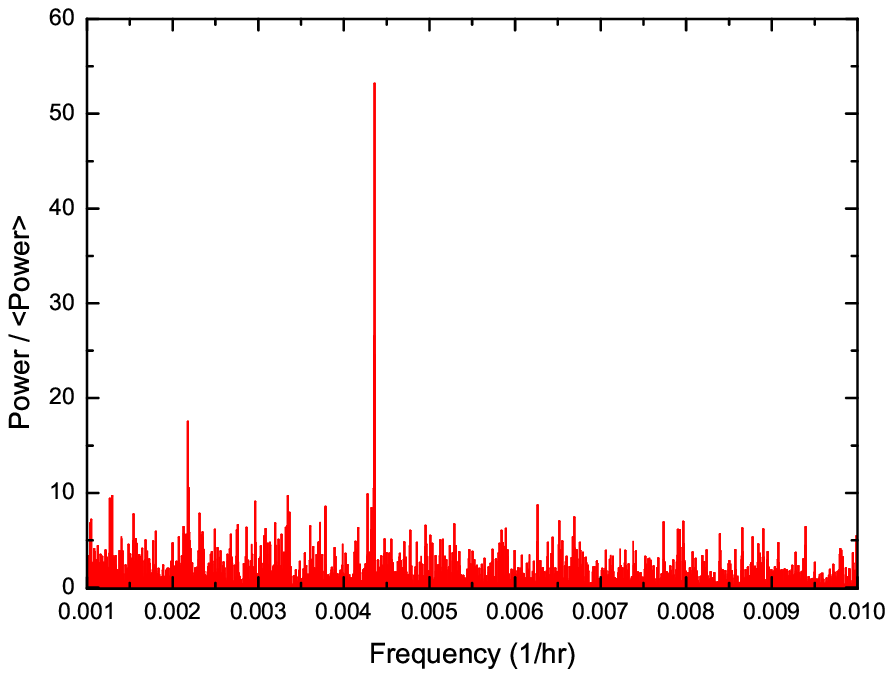}
\caption{Power spectra of the RXTE/ASM light curve of 4U 2206+54
with observations from 1997 to 2008. The left panel shows the
power spectrum in all frequency band. And the right panel presents
the zoom-in of the lower frequency band which clearly shows the
position of two peaks at $\sim 9.56$ days and 19.11 days. }
\end{figure*}

\begin{figure*}
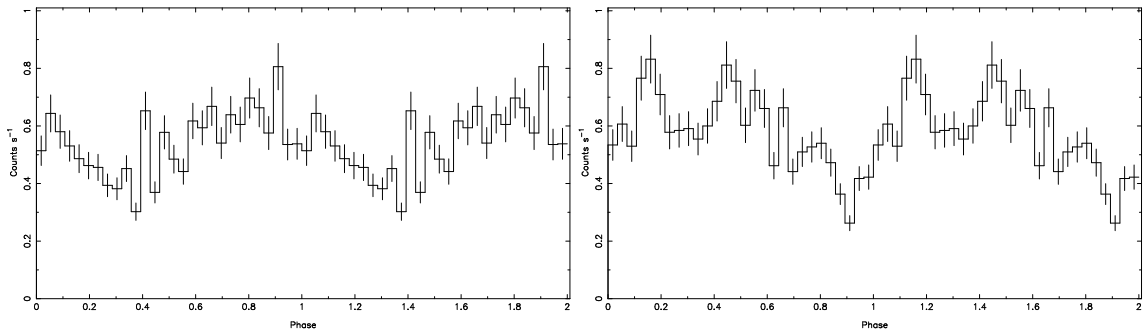

\centering
\includegraphics[angle=0,width=7.5cm]{2206_9d.ps}
\includegraphics[angle=0,width=7.5cm]{2206_19d.ps}
\caption{RXTE/ASM light curves of 4U 2206+54 folded at two
possible orbit periods at 9.56 day (left) and at 19.11 day
(right). }
\end{figure*}

From the ASM light curve of 4U 2206+54, we found two modulation
periods of $\sim$ 9.56 days and the twice, 19.11 days. Then which
one is the real orbit period in 4U 2206+54? One possibility is
that the 19.11-day period is just the first harmonic of the
9.6-day period. But the SWIFT/BAT data only found one period at
19.2 days, and recent RXTE/ASM data from 2004 -- 2006 showed the
modulation period at 19.1 days, not significantly at 9.6 days yet
(Corbet et al. 2007; we also folded the light curve at 9.6 day
using the RXTE/ASM data after 2003, no significant features were
detected in phase profiles). Then the 9.6 day period was not a
permanent signal in X-ray light curve. Then appearance of 19.2-day
period may not be the first harmonic of 9.6 day. In addition,
optical radial velocity measurements (Blay et al. 2006) have not
shown the period at 9.6 days but the power spectrum of the
complete set of radial velocity observations suggested a peak at
$\sim 14.89$ days, which was interpreted as an artifact.

Modulation of X-ray emission on the orbit period in a high mass
X-ray binary may occur in different ways as follows. Firstly, if
the orbit is eccentric, the peak emission would occur at
periastron passage ($e=0.15$ assumed by Rib\'o et al. 2006). In
this case, it is difficult to understand why two periods appeared
in the X-ray light curve. In addition, an eccentricity of $\sim
0.4$ was suggested from the X-ray light curve analysis (Reig et
al. 2009), so 4U 2206+54 has a modestly eccentric orbit. In the
second case, if the orbit plane is inclined to the plane of an
equational circumstellar disc, two possible outbursts could occur
when the neutron star passes through the disc. This scenario has
been proposed to explain two modulation periods detected in two
X-ray binaries: a Be/neutron star binary GRO J2058+42 (Corbet et
al. 1997) and a supergiant system 4U 1907+09 (Mashall \& Ricketts
1980). Though 4U 2206+54 has the difference in the nature of the
mass donor, the orbit behavior may be similar to the case in GRO
J2058+42 and 4U 1907+09. Therefore, we suggested that the orbit
period of 4U 2206+54 should be $\sim$ 19.11 days, which was
consistent with the conclusion by Corbet et al. (2007).

In hard X-ray band, only the orbit period at $\sim 19$ days could
be found with SWIFT/BAT observations (Corbet et al. 2007), while
in soft X-ray band of 1.5 -- 12 keV, we found two modulation
periods. If the 9.6-day period is due to the neutron star passing
through the equational plane of the mass donor, soft X-ray
emission may be affected by the enhanced wind around the disc, but
hard X-ray emission has no significant dependence on it.
Therefore, the hard X-ray emission above 20 keV and its variation
pattern should have a different origin from those in the soft
X-ray band, which requires more studies.

In the Be star/neutron star binary systems, there exists a
possible correlation between orbit period and pulse period (Corbet
1986). Though 4U 2206+54 have a different mass donor but may be
still similar to a Be star system, it is predicted to have a pulse
period of a few second for a orbit period of 19.11 days according
to the correlation, which however is unconsistent with the
pulsation period of $\sim$ 5400 s reported in this work (maybe
$\sim$ 5560 s, also see Reig et al. 2009). Two high mass X-ray
binary systems also show the similar behavior away from the
correlation: A0538-66 with an orbit period of 16.7 days has an
short pulse period of 69 ms (Skinner et al. 1982) and SAX
J2103.5+4545 with an orbit period of 12.7 days has a longer pulse
period of 358.6 s (Baykal et al. 2000). Generally, the 19 day
orbit period is relatively short for a Be star system. Therefore,
the orbit-pulse period correlation would be unvalid at the short
orbit period. From pulse period versus orbit period diagram (see
Reig et al. 2009), 4U 2206+54 fall into the wind-fed supergiant
system region which was consistent with the proposed
wind-accretion scenario from the X-ray emission properties (Corbet
\& Peele 2001; Rib\'o et al. 2006).

Though we have now confirmed the nature of the compact object in
4U 2206+54, many questions on this high mass X-ray binary are
still unclear, requiring further studies in both theories and
observations. Origin of the low-pulsation neutron star in 4U
2206+54 is a mystery according to the present scenario of the
neutron star's spin evolution in a close binary system (Davies \&
Pringle 1981; Li \& van den Heuvel 1999). So Reig et al. (2009)
suggested that the neutron star in 4U 2206+54 was born as a
magnetar with $B\geq 10^{14}$ G which was used to explain the long
pulsation period of 2S 0114+65 (Li \& van den Heuvel 1999). Since
the donor star of 4U 2206+54 is a main sequence massive star, the
magnetic field of the neutron star decaying from $\geq 10^{14}$ G
to 10$^{12}$ G during wind accretion processes within several
million years is also questionable. There exist other possible
ways to resolve the problems, like X-ray pulsar in 4U 2206+54 may
not follow the present standard evolution models in close
binaries. Some numerical simulations suggested that no significant
angular momentum transfers onto the neutron star from the wind of
supergiant systems (Ruffert 1999). In the case of 4U 2206+54 with
a magnetic neutron star, it is possible that the calculations
(Ruffert 1999) were inapplicable, so that the 5000-s pulsation
period was formed during the wind-fed accretion phase.

We discovered cyclotron resonant absorption lines at $\sim 30$ keV
60 keV in the active state of 4U 2206+54, but no detection of
cyclotron lines in the quiescent state. In previous studies, the
possible detection of electron cyclotron absorption line at $\sim
30$ keV was reported by observations of RXTE, BepposSAX and
INTEGRAL(Torrejon et al. 2004; Masetti et al. 2004; Blay et al.
2005). But recent RXTE observations by Reig et al. (2009) did not
find the cyclotron line at $\sim 30$ keV similar to the quiescent
state reported in this paper. Then from the present data,
cyclotron lines could be detected when the hard X-ray luminosity
(20-- 100 keV) is higher than $10^{35}$ erg s$^{-1}$ (also see
Blay et al. 2005). We suggested that in the active state (like
$L_{\rm 20 - 100 keV}>10^{35}$ erg s$^{-1}$), hard X-ray emissions
will mainly come from the surface (e.g., polar cap region) of the
neutron star, while in the quiescent state, the X-ray emission
would mainly come from the standing point region where the
magnetic pressure is equal to the accretion pressure.

Among the wind-fed massive binary system, 4U 2206+54 is a unique
one which was identified as a magnetic neutron star with a very
low pulsation. Thus, detailed studies and long-term monitoring of
the peculiar neutron star binary 4U 2206+54 in X-rays will help to
understand accretion physics and evolution of wind-fed systems.

\section*{Acknowledgments}
The author is grateful to the referee for the fruitful suggestions
to improve the manuscript and also to Jiang, Y. Y. for the help to
prepare some figures. This paper is based on observations of
INTEGRAL, an ESA project with instrument and science data centre
funded by ESA member states (principle investigator countries:
Demark, France, Germany, Italy, Switzerland and Spain), the Czech
Republic and Poland, and with participation of Russia and US. We
used the archival data from ASM Light Curve webpage developed by
the ASM team at the Kavli Institute for Astrophysics and Space
Research at the Massachusetts Institute of Technology. W. Wang is
supported by the National Natural Science Foundation of China
under grants 10803009, 10833003.

\end{document}